\documentclass[sigconf]{acmart}

\usepackage[capitalise]{cleveref}

\begin{document}

\title{It's Complicated: On the Design and Evaluation of AI-Powered AAC Interfaces}

\author{Blade Frisch}
\email{bwfrisch@mtu.edu}
\affiliation{
  \institution{Michigan Technological University}
  \city{Houghton}
  \state{Michigan}
  \country{USA}
}

\author{Will Wade}
\email{will.wade@thinksmartbox.com}
\affiliation{
  \institution{Smartbox Assistive Technology Ltd}
  \city{Bristol}
  \country{UK}
}

\author{Dylan Gaines}
\email{dgaine20@kennesaw.edu}
\affiliation{
  \institution{Kennesaw State University}
  \city{Marietta}
  \state{Georgia}
  \country{USA}
}

\author{Michelle Kinsella}
\email{kinsella@ohsu.edu}
\affiliation{
  \institution{Oregon Health \& Science University}
  \city{Portland}
  \state{Oregon}
  \country{USA}
}

\author{Betts Peters}
\email{petersbe@ohsu.edu}
\affiliation{
  \institution{Oregon Health \& Science University}
  \city{Portland}
  \state{Oregon}
  \country{USA}
}

\author{Tamara Broderick}
\email{tbroderick@mit.edu}
\affiliation{
  \institution{Massachusetts Institute of Technology}
  \city{Cambridge}
  \state{Massachusetts}
  \country{USA}
}

\author{Keith Vertanen}
\email{vertanen@mtu.edu}
\affiliation{
  \institution{Michigan Technological University}
  \city{Houghton}
  \state{Michigan}
  \country{USA}
}

\acmConference[Speech AI for All Workshop at CHI]{Speech AI for All: The What, How, and Who of Measurement Workshop at the CHI Conference on Human Factors in Computing Systems}{April 16, 2026}{Barcelona, Spain}
\acmDOI{}
\acmISBN{}
\copyrightyear{2026}
\acmYear{2026}
\setcopyright{rightsretained}

\renewcommand{\shortauthors}{Frisch et al.}

\begin{abstract}
    Artificial intelligence (AI) can enhance what people who use augmentative and alternative communication (AAC) are able to do with their systems. However, evaluating AI-powered AAC interfaces can be difficult. People are intersectional beings and current evaluation metrics can struggle to capture the multifaceted and nuanced desires people may have for their AAC. We explore the complicated nature of six AAC problem spaces, explore how AI might be used in these spaces, and suggest more robust methods of evaluation that take the intersectional nuances of people into account. We also discuss broader issues that arise across these problem spaces and how they could be addressed using our proposed evaluation methods.
\end{abstract}



\maketitle

\section{Introduction}\label{introduction}

Metrics are essential for evaluating software performance and user experience. In order to determine how well a piece of software is performing, we must find a way to measure that performance. In machine learning and natural language processing, quantitative benchmarks are the standard for validating model efficacy. However, when these and other artificial intelligence (AI)-based approaches are integrated into augmentative and alternative communication (AAC) systems, traditional efficiency metrics often fail to capture the complexity of a user's requirements. Evaluating AI-powered AAC requires a combination of technical performance metrics and human-centric data (e.g., self-reported usability measurements~\cite{albert_measuring_2023}, interviews~\cite{seidman_interviewing_2019}, usability testing~\cite{barnum_usability_2021}).

We argue that the evaluation of AI-powered AAC interfaces must move beyond the assumption that a user's needs and abilities are static and singular. We also argue that these evaluations should not be governed solely by technical requirements and must include the needs and desires of AAC users. While technical performance metrics are necessary for system development, any finite set of technical evaluations is inherently incomplete. We propose that a richer, pluralistic set of evaluation methods can better capture misfits between an AAC user and their communication context. Our primary contributions are:
\begin{itemize}
    \item \textbf{Six AAC design considerations ---} We analyze six design considerations to take into account when evaluating AAC interfaces: speed and accuracy, mental and physical effort, agency in identity presentation, adapting to communication contexts, turn-taking, and fluctuating physical ability.

    \item \textbf{Possible AI-powered features for AAC ---} We propose various ways AI might be leveraged to improve the efficacy and user satisfaction with AAC systems.
    
    \item \textbf{Multidimensional evaluation methods ---} We propose evaluation methods that combine technical system metrics with human-centered design research methods to capture a broader understanding of AAC users' needs.

\end{itemize}

\section{Positioning}

Existing research on AI in assistive technology is often bifurcated. Research often focuses on optimizing predictive models for speed~\cite{koester2018text}. While technical performance metrics are useful for evaluating a model's performance, they may not be measuring the whole story of what matters to AAC users. If they are the sole method used to evaluate a model, they can impose artificial limitations on AAC systems solely to satisfy technical requirements rather than to ensure the system is addressing what truly matters to AAC users. Such a focus is similar to the medical model of disability, which risks pathologizing a communication disability over supporting the unique identities and desires of disabled people~\cite{kafer_feminist_2013, withers_disability_2012}. We must move beyond imposing solutions on AAC users and instead work with them to create new approaches.

It is critical to engage with AAC users in the design and intervention process \cite{light_putting_2013}. This should include learning about what is important to AAC users and how to measure an AAC system's ability to support those priorities, rather than selecting measurements because they are convenient from a technical perspective. AAC users and advocates have cautioned against viewing AI and other technological advances as ``cures'' for disability~\cite{sellwood2024imagining}. While recent work has explored AI bias in domains like facial recognition~\cite{buolamwini_unmasking_2024} and negative outcomes that can come from introducing AI-powered systems~\cite{venkatasubramanian_toward_2025}, more work is needed to explore these issues in the AAC domain.

This paper bridges these perspectives. We examine the validity of standard technical performance metrics through the lens of intersectionality and argue that AAC lacks a unified evaluation strategy that accounts for the dynamic nature of communication or the diverse needs and preferences of users. We propose evaluation methods that combine AAC users' wants and needs with technical performance metrics.

\section{Related Work}\label{related_work}

AAC users are not one easily-defined group. There are many different identities people have (e.g., race, gender, societal role, views, disability status), and a person exists at the intersection of these identities~\cite{tracy_qualitative_2020}. This intersectionality can influence how they interact with technology. Problems arise when disability is reduced to only one identity; instead, disability can, and should, be viewed as intersectional~\cite{kafer_feminist_2013, withers_disability_2012}. Rosemarie Garland-Thomson uses this lens to describe disability as a ``misfit'': a relational encounter between an individual's body-mind and an unaccommodating environment~\cite{garlandrosemary}.

AI models often attempt to simplify this intersectionality by optimizing for a single target identity. As demonstrated by \citeauthor{buolamwini_unmasking_2024} \cite{buolamwini_unmasking_2024}, such simplification can cause models to perform poorly for users at the intersection of marginalized identities \cite{buolamwini_unmasking_2024}. In the context of AAC, this leads to technoableism \cite{shew_against_2023}, where systems are optimized for a hypothetical standard user and ignore that disability is intersectional and that individuals have multiple identities. Technical performance metrics can provide useful data about individual components of an interface or model, but they must be part of a larger suite of tools, methods, and measurements to better capture the intersectional nature of the people who use AAC and AI.

AAC users desire agency in the communication process, retaining control over who they interact with and how they communicate with people \cite{frisch_designing_2025}. People change their behavior based on a myriad of social factors \cite{goffman_presentation_1959}, and being disabled can lead to being stigmatized in social situations \cite{goffman_stigma_1963}. Designing AAC interfaces without supporting the agency of AAC users to control their communication and their presentation of self can impact how people use AAC and how others perceive them.

When designing AAC interfaces, it may be important to consider how users value their ability to respond in a timely manner during conversations to follow social norms of communication~\cite{american_speech-language-hearing_association_components_nodate}. By definition, communication involves more than just one person. AAC users will be communicating with other people, and the characteristics of these conversations may affect how AAC users and their AAC systems are perceived by others~\cite{hoag_hierarchy_2008}. Communication partners will bring pre-existing beliefs and biases to conversations with an AAC user. The partners' opinions may be further shaped by the AAC strategies and access methods used by the AAC user~\cite{kane_at_2017}. These perceptions can affect the success of communication interactions and the effective participation of AAC users in a variety of life situations~\cite{peters2023effects}. Familiar communication partners may also be a valuable source of personally- or situationally-relevant information, and some AAC users may desire the ability to leverage the real-time input of these partners to supplement existing word prediction models while maintaining the autonomy to make the final decision about word or phrase selection~\cite{Fried-Oken02102025smartpredict,Roark02012015_huffman_linear_scanning}.

AI-powered features have the potential to influence partner perceptions of AAC users positively (e.g., by supporting the user in producing messages that conform to the partner's expectations for the conversation) or negatively (e.g., by giving the erroneous impression that message content is being controlled by the AI and not by the user). Understanding the perspectives of communication partners could help inform best practices for training and educating those partners about what AI can and cannot do, managing expectations, and encouraging partners to presume communicative competence.

\section{Exploring the Who, What, and How of AAC Interface Design}\label{problems}

This section explores six critical domains where AI can augment AAC. However, as established in \cref{related_work}, every technical intervention risks othering the user if evaluated through a narrow lens. Our goal is to move beyond the hypothetical standard user fallacy by proposing evaluation methods that account for individual identity and environmental misfits.

\subsection{Communication Speed and Accuracy}\label{problem_01}

Spoken communication can be very fast at over 150 words per minute (WPM). Text-based AAC interfaces can be substantially slower (e.g.,~2 WPM \cite{koester2018text}), especially for users who require alternative access methods or interfaces due to physical or sensory impairments. This large disparity in speech production rates can be disruptive to communication~\cite{alm_prediction} and even lead communication partners to form negative perceptions about AAC users~\cite{beck2000attitudes}. The speed of text production is not the only factor of communication to consider; users also want to accurately communicate their thoughts, and different users will have different preferences concerning the speed-accuracy trade-off \cite{fried2024stakeholders}. The accuracy of AAC output can be affected not only by input errors but also by incorrect inferences made by any AI that an AAC system might use. 

\subsubsection{What AI Might Do}

AAC interfaces often predict a user's upcoming text based on their previous text~\cite{garay_survey}. These interfaces may make letter predictions (e.g.,~\cite{ward_acm,roark2010scanning}) or word predictions (e.g.,~\cite{trnka_user,wandmacher_sibylle}). Non-literate users might use phoneme prediction (e.g.,~\cite{trinh2010further,trinh_iscan}) or symbol prediction (e.g.,~\cite{gatti2006caba2l,stewart2000improving}). Historically, such letter, word, phoneme, or symbol predictions have leveraged statistical n-gram language models. Recent results show that large language models (LLMs) based on neural networks can improve letter and word predictions \cite{gaines_aacllm}. Given LLMs' ability to transfer their knowledge to new languages despite limited training data~\cite{loreslm_lowresource}, they likely can offer improved predictions for phonetic and symbolic AAC systems as well. 

While predicting the next letter or word can be helpful, the ``holy grail'' of AAC interfaces has long been to predict entire utterances to speed up communication (e.g.,~\cite{demasco1992generating,todman2008whole}). 
The real promise of LLMs is that they may be able to generate ``big swing'' predictions that predict multiple words or entire sentences. LLMs can not only leverage substantially more of a user's previous writing, they can also integrate other contextual clues. For example, AI might glean such clues from listening to the conversations around the user~\cite{wisenburn2008aac,adhikary_speech}, integrating suggestions from their communication partner~\cite{roark2011towards}, using a camera to infer things about a user's surrounding environment~\cite{kane2017let}, or identifying who the user is speaking with~\cite{kane_talkabout}. Whether AAC users or the people around them would want such contextual monitoring is a topic worth future investigation.

\subsubsection{Possible Ways To Evaluate}

While metrics such as words per minute are commonly used to measure communication speed, they must also be balanced with some way of measuring accuracy (i.e.,~it is no good producing text quickly if it is not what the user wanted to say). Accuracy might be measured in different ways depending on the situation; for example, in some contexts, specific wording may be very important to the user, while in other contexts it may only matter if AI-generated text captures the essence of what the user wanted to say. Because the correctness of an utterance is often subjective and context-dependent, we propose a multidimensional approach:

\begin{itemize}
    \item \textbf{Semantic and linguistic similarity (offline) ---} While traditional metrics like word error rate (WER) measure verbatim accuracy, they fail to account for AI predictions that capture the meaning but not the exact phrasing. This might be possible by utilizing LLM judges and semantic embeddings that better assess intent preservation than simple text matching metrics such as WER.

    \item \textbf{User-mediated assessment (online) ---} In interactive user studies, accuracy should be measured through the user's perception of how well the produced text aligns with their intent. 
    This includes tracking correction rates, noting when a user accepts a close-enough prediction versus when they feel compelled to edit it. This can reflect the pragmatic trade-off the user may make between speed and effort.

    \item \textbf{Functional success (task-based) ---} Evaluation can be shifted from the text itself to the outcome of the social interaction. By using task-based scenarios, researchers can measure whether the user's communication accomplished the task, such as communicating a specific idea or requesting an action. This can validate the system's efficacy as a tool for participation rather than just a text entry interface.
\end{itemize}

Measuring the user's intent remains a significant methodological challenge. While lab settings often provide participants with pre-defined communicative goals, longitudinal contextual evaluations could have users review their own logs to rate the accuracy of AI-assisted utterances and how much agency they retained when composing text.

In lab user studies, it is also common to measure entry and error rate by having participants copy fixed phrases. This provides an easy way to measure error rate, but differs from how people use AAC systems in practice (i.e.,~converting their own thoughts into text). While it makes error rate measurement more difficult, it is possible to allow participants to freely compose messages~\cite{vertanen_complementing}. In the case of both text copy and free composition tasks, it is important to ensure an AAC system can input whatever a user desires, including difficult-to-predict words (e.g.,~proper names, acronyms, passwords). One approach to measuring this aspect of an AAC system is to have participants copy phrases with rare words~\cite{vertanen_velociwatch} or ask participants to compose text they suspect will be difficult for the system~\cite{gaines_enhancing}. 

\subsection{Physical and Mental Effort}\label{problem_01b}

Some people with communication disabilities also have physical disabilities that make it difficult or impossible to use standard text entry devices, such as a keyboard or touchscreen. Instead, they may use alternative access methods that allow them to compose text using other interaction methods, such as eye gaze or switches. People who use AAC, and particularly those who use alternative access methods, may experience fatigue, eye strain, or other physical symptoms as a result of system use. Some access methods, such as switches, make it possible to control a computer with minimal movement but may require more than one user action per selection. For example, in the single-switch AAC system Nomon~\cite{bonaker_nomon,bonaker_usability}, the user may need to activate a switch multiple times before the system is confident enough to select a specific character or word. In such systems, it becomes crucial to consider the required physical effort to enter text and balance it with other factors such as speed and accuracy.

It also takes mental effort to process partner communications, navigate and use the AAC interface, and choose between future actions sequences (e.g.,~how to correct any previous errors). Correcting errors can be frustrating \cite{alharbi2022autocucumber} and dominate a user's time \cite{azenkot-exploring}. Some access methods, such as brain-computer interfaces or eye-tracking, require constant mental attention, with limited opportunities for the user to take a break. When an AAC interface requires several taxing actions, the user can be left both physically and mentally exhausted.

\subsubsection{What AI Might Do}

An AI may be able to detect when a user is experiencing increased physical or mental effort. This might be done via physiological sensing (e.g.,~over-the-ear EEG sensors integrated into smartglasses) or by observing a user's previous interactions. This information could be used to dynamically adapt the interface to reduce effort. For example, it may be better to use a more expensive language model that offers more accurate predictions, even if the added prediction latency slows input. 

Another option might be to increase the number of predictions displayed. This may increase the mental effort required to search through the presented predictions, but if the predictions are accurate, it could reduce the physical effort required to input that text. However, this may depend on a user's particular preferences, abilities, and access method (e.g.,~in Nomon more prediction targets can increase the physical actions required per selection). 

We previously discussed the idea of an AI making ``big swing'' predictions (i.e., predicting multiple words or sentences at once). As a user becomes fatigued, they may increasingly accept such bigger predictions and the system could increasingly present them. While we already discussed how this can increase entry rates, it can also reduce the number of input actions required by the user to enter text, especially in interfaces such as Nomon and RSVP~\cite{orhan-rsvp} where multiple user actions are required to make a selection. 

If the predictions are not accurate, it can increase the number of user actions required (and therefore the effort required) to input text or correct errors made by the system. The AI could dynamically detect when the user is engaging in more corrective actions and reduce the weight of the language model to prevent additional AI-induced errors. The system could also potentially detect a user's mental response to an erroneous selection via error-related potentials~\cite{yasemin-errp} sensed via EEG and engage in automatic error correction.

\subsubsection{Possible Ways To Evaluate}

Quantifying the physical effort required for system use can depend on the specific interface and access method. For example, in Nomon, the authors measured how many switch activations (e.g., pressing a button, blinking using eye-gaze tracking, releasing a puff of air in a sip/puff switch) were required to select a target. Another option is the CARE Efficiency Score \cite{AACEffortAlgorithms} which estimates the level of effort required to activate buttons based on motor distance and visual scanning. This allows us to evaluate if AI's big swing predictions actually reduce the motor load or if the cognitive and physical costs of correcting a major error negate the potential gains. 

To quantify mental effort, it can be beneficial to assess user workload with self-reported questionnaires such as the NASA Task Load Index \cite{hart1988development} or variations adapted for human-computer interaction and assistive technology applications~\cite{bates2006enhancing,peters2016soliciting}. It may also be possible to use sensor-based techniques to estimate mental effort, such as by using eye-tracking~\cite{chen_eye_mental} or EEG~\cite{zhu-mental-effort}.

\subsection{Sounding Like You Want to Sound}\label{problem_02}

Speech-generating AAC can often fail to correctly express voice tones, accents, cadences, and other aspects of speech that shape how people communicate \cite{kane_at_2017}. \citeauthor{Judge2013-qw} \cite{Judge2013-qw} found that users often prioritize vocal quality and regional accents as a means of maintaining social presence. The inability of an AAC system to accurately replicate the nuances of natural speech can lead to AAC users being misunderstood and misrepresented by their means of communication, and may affect communication partners' perception of the user.

\subsubsection{What AI Might Do}

AI can be used to generate a voice model for a person based on banked voice data \cite{Chen02102023}. If there is no such data available, AI could also be used to synthesize a voice based on existing speech samples from other people. This could even include blending different voice samples to achieve specific attributes, such as creating a specific accent. AI could also change aspects of the voice generated by text-to-speech based on tone indicators inserted by the user and by context clues, such as what and how the communication partner is communicating. These tone indicators could instruct the text-to-speech engine on where to include pauses or how to change the pitch of a word or phrase. It could also learn the tone indicators typically selected by the user for specific phrases or frequent conversation topics and suggest them when composing text, learning their unique communication style over time.

\subsubsection{Possible Ways To Evaluate}

Usability testing can be used to evaluate the tone indicators and the user's satisfaction with both the indicators and the generated speech. Diary studies (i.e., where a person will use a piece of software for a length of time and write diary entries about their experiences using the software) can then be used to see how AAC users integrate these tools into their daily communication lives. We can also measure the impact of different tone indicators by conducting usability testing on each tone indicator individually and collecting data on user satisfaction with the indicator under test.

Evaluation could include rating scales where users rate the self-identification of the voice, assessing whether the output feels like an extension of themselves or not. Additionally, gathering satisfaction data from both AAC users and their communication partners on the generated speech can help determine if the AI-generated prosody successfully conveys the user's intended emotion or social standing to their conversation partners \cite{kane_at_2017}.

\subsection{Code- and Context-Switching}\label{problem_03}

People will change how they communicate based on the details of the situation. They may switch between languages or dialects when communicating, which is called code-switching \cite{van_herk_what_2018}. They may change other aspects of their communication (e.g., usage of slang, level of formality) based on context: who they're communicating with, the number of communication partners, and the content being communicated. The environmental and social context surrounding the communication also matters: conversing in a loud coffee shop and sharing information in a doctor's office have different communication needs and present different challenges to AAC users \cite{peters2023effects,baylor2013communicative}. Current AAC systems offer limited support for users in code- and context-switching to better adapt to different communication partners and situations (e.g., chatting with a friend, participating in a job interview, or sharing current symptoms with a doctor), or in switching between languages for multilingual users \cite{kane_at_2017}.

\subsubsection{What AI Might Do}

AI systems can be trained to change their predictions based on who the AAC user is communicating with and the context in which the user is communicating. For example, there could be a mode to predict more friendly and casual text for chatting with friends, where the predictions are more concerned with the user sounding like themselves \cite{Klein02042024} (see \cref{problem_02}). Another mode might focus on the user's formal, professional tone for contexts such as job interviews. Still another mode would be helpful when communicating with doctors, where predictions should focus more on accuracy in communication and helping the AAC user say exactly what they need to communicate to their doctor. AI could also detect when utterances, or portions of an utterance, are in a different language and adjust the pronunciation of the words based on the indicated language.

\subsubsection{Possible Ways To Evaluate}

Evaluating code- and context-switching first requires understanding the different communication styles and contexts AAC users have. This can be done through semi-structured interviews, contextual inquiry, ethnography, and questionnaires like the Communication Needs Questionnaire \cite{bardach2017communication}. Once the styles and contexts are understood, the success of the code switch can then be measured through methods like usability testing and user satisfaction measurements. The Communicative Participation Item Bank (CPIB) \cite{baylor2013communicative}, a self-reported measure of the difficulty an individual experiences when participating in various communication-related situations, may also be useful for evaluating the success of code-switching to adapt to different communication contexts.

\subsection{Taking Part Fluidly in a Conversation}\label{problem_04}

Turn-taking is one aspect of interpersonal interactions where AAC users can struggle to communicate \cite{peters2023effects}. It takes significantly more time to compose messages on AAC devices than it does to speak them, often leading to the conversation moving on or changing topics before the user can contribute \cite{kane_at_2017}. \citeauthor{Judge2013-qw} \cite{Judge2013-qw} found that users identify the ability to be fast and spontaneous as a key design requirement, yet one that is frequently unmet, leading to a restricted use of the device in dynamic social settings. \citeauthor{weinberg_one_2025} \cite{weinberg_one_2025} show that AAC users can also struggle to make use of backchanneling (e.g., adding interjections like ``hmm'', ``yeah'', and ``uh-huh'') when interacting with both other AAC users and non-AAC users. This can also include when AAC users try to capture the attention of their communication partners in order to ``take the floor'' and become the active speaker~\cite{valencia_aided_2021}.

\subsubsection{What AI Might Do}

AI can predict the backchanneling methods that the participant would like to use in the conversation and adjust the AAC system to make use of those methods. Additionally, the AAC system could include more efficient ways to predict take-the-floor utterances, such as ``Can I chime in?'' or ``Oh yeah, that reminds me of something!''. Scripting interactions is a strategy that some AAC users make use of to prepare for social and community-based interactions. AI can be used to help prepare scripts for these interactions so that utterances are readily available for an AAC user to use when communicating with others.

\subsubsection{Possible Ways To Evaluate}

Success in turn-taking cannot be measured by text accuracy alone; it can also be measured by latency to respond and floor-taking success rates. Quantitatively, we can measure the time difference between when the user wants to take the floor and when the take-the-floor signal occurs. However, because communication is a relational act, this could be complemented by interviewing communication partners. These interviews could reveal whether the AI-guided backchanneling and floor-taking make the user appear more present in the exchange, or if the AI's timing feels uncanny or inorganic to the partner. This will help guide AAC designers in ensuring the AAC user retains their agency in self-presentation. The CPIB includes items directly related to fast-moving conversations, communication in group settings, and other contexts involving backchanneling and turn-taking, and could be used to assess the impact of AI-based features on participation in those situations. This dual approach ensures that while the AI speeds up the mechanics of turn-taking, it maintains the authenticity of the user's social presence.

\subsection{Short- and Long-Term Needs Changes}\label{problem_05}

Communication needs can be dynamic and change on short- and long-term scales. For example, autistic people can have an increased need for alternative methods of communication as they become overstimulated, which decreases as the stimuli are removed. People with ALS may experience short-term changes due to fatigue or the effects of medication, but because of the degenerative nature of the condition, there can be changes in physical function over a longer period of time. These changes in physical ability can change how people interact with their AAC system, such as changing from using a touch screen to using an eye tracker or brain-computer interface. 

A static AAC system might not be able to support short-term changes or be usable in the long-term as an individual's disability progresses. This can create an additional burden on the user to continuously learn and adapt to a new system when their old system is no longer able to support their needs. In addition to accommodating users' physical, cognitive, or emotional states, AAC systems should also be able to adapt to their changing communication needs and preferences. An individual will encounter different communication contexts, partners, and topics as they move through life, and will want to express themselves in different ways.

\subsubsection{What AI Might Do}

AAC systems can be designed with multiple interface options based on the different communication needs a person may have. For example, an AAC system designed to support autistic people who experience shutdown could have two ways of interacting with composed utterances: an interface that provides fine-grained controls for text-to-speech (e.g., speech rate controls, speaking only portions of the utterance at a time) and a simplified interface that relies on preset controls to speak an entire utterance. AI can then learn the user's behavior and change from the detailed interface to the simplified interface if it detects that the autistic person is entering shutdown and needs less stimulation.

It can also adjust its text predictions based on the detected communication needs, such as offering larger predictions when it detects that the user needs more communication support. As discussed in \cref{problem_01,problem_01b}, these ``big swing'' predictions can help increase communication speed and reduce the physical and mental effort of composing longer utterances. The size of these predictions and how the predictions are made could be adjusted based on the user's needs. For example, people with ALS may prefer to make less use of predicted text when their physical function is relatively strong (e.g., earlier in disease progression or earlier in the day). But they may benefit from additional predictions, including entire sentences or multiple sentences, when they are more fatigued or when they are using slower alternative access methods.

\subsubsection{Possible Ways To Evaluate}

Evaluating changing needs requires a longitudinal approach. While it may be necessary to use technical metrics to evaluate an AI model's ability to detect the current needs of an AAC user, these cannot wholly capture how AAC users would respond to an adaptive system. Usability testing can show the initial reactions of AAC users to a dynamic interface, but it can be difficult to detect changes in an AAC user's communication needs on the time scale of a usability test. Diary studies are a useful tool for measuring longitudinal data, which can help measure needs changes over time. An AAC user could be given an adaptive AAC system and be asked to write diary entries whenever the system detects a needs change and adapts the interface. The AAC user could then provide information on how the adaptation impacted their communication and whether they felt the adaptation should have been made. Combining these qualitative data with technical metrics on needs detection tells a more complete story of how an AI model can adapt an AAC interface based on changes in user needs.

\section{Discussion and Limitations}\label{discussion}

The individual design challenges explored in \cref{problems} are underpinned by broader, systemic issues. As noted by \citeauthor{Bennett2023-bj} \cite{Bennett2023-bj}, designing data-driven technologies for long-term conditions often risks epistemic injustice, where technical performance metrics are preferred over a user's lived experience. In AAC, a high text entry rate can mask a user's loss of authentic voice or autonomy. This creates an issue: an AI system may be successful by the developer's metrics while failing the user's sense of self. \citeauthor{Konadl2024-vc} \cite{Konadl2024-vc} highlights that current generative solutions often fail to bridge the gap between formal and informal contexts, forcing one-size-fits-all output. This creates a misfit between an AAC user's needs and what the AAC system supports \cite{garlandrosemary}, as the user is coerced into normative speech patterns to satisfy the model's requirements. This can devalue their identity in favor of technically-driven evaluation.

To address this issue, AAC evaluation could be situated within the Three Domains Framework proposed by \citeauthor{Judge2013-qw} \cite{Judge2013-qw}. While AI development often focuses solely on the ``Device Design'' domain (speed and reliability), successful use is equally dependent on the ``Wider Picture'' (environmental support) and the ``Personal Context'' (user identity) domains. Clinical tools, such as the Individually Prioritised Problem Assessment (IPPA) \cite{wessels2002ippa} and the Communicative Participation Item Bank \cite{baylor2013communicative}, can offer a potential starting point for uncovering the requirements across these three domains. By moving beyond simple satisfaction scores, these person-centered frameworks can help determine if AI features are fostering genuine communication growth.

Furthermore, as interfaces become more proactive, offering AI-guided changes based on perceived context or fatigue, we face a significant UX challenge: providing granular autonomy within an interface that is already inherently difficult to access. As \citeauthor{Judge2013-qw} \cite{Judge2013-qw} demonstrate, users identify a fundamental link between communication speed and the dignity of the interaction. However, they also warn that simplicity of design remains a key requirement. AI interventions intended to enhance communication must not inadvertently increase the cognitive load or decrease system reliability.

We propose several potential paths for how AI can be introduced into AAC interfaces in \cref{problems}. However, these potential implementations have not yet been validated or tested with AAC users. As we discuss throughout this work, evaluating AI in AAC interfaces is complicated and should not be done with quantitative metrics alone. Any inclusion of AI in AAC interfaces should be done alongside AAC users via participatory design approaches. This will help ensure that the desires and needs of AAC users are included when evaluating an AI-powered AAC interface.

\section{Conclusion}

Technical performance metrics alone can struggle to collect intersectional data. It is nearly impossible to capture the many dimensions of one's identity solely from a single, or even a set of, technical performance metrics. Instead, evaluating an AI-powered AAC interface must be done with a pluralistic set of evaluation methods guided by the wants and needs of AAC users. Both technical performance metrics and human-centered data are needed to tell the entire story of how an interface is performing. It is also critical to ensure that what is measured is guided by what AAC users decide is important. Trying to collapse people into a single identity and collect only technical performance metrics from this narrowed viewpoint is a path that leads to technoableism.

Researching humans is complicated. We have explored this complexity across six AAC problem spaces and posited that people do not fit into neat boxes. Technical performance metrics for evaluating machine learning and natural language processing models can often fail to capture all of what AAC users value in their AAC systems. Just as there is great nuance in understanding these technical performance metrics and applying them correctly, there is nuance in connecting metrics to the broader user experience. Technical performance metrics must be only one star in the constellation of evaluation so that the entire story of a user's experience with their AAC system can be told.

\begin{acks}
    This work was funded in part by the National Science Foundation (IIS-2402876, IIS-2402877, and IIS-2402878).
\end{acks}

\balance

\bibliographystyle{ACM-Reference-Format}
\bibliography{references}

\end{document}